\documentclass[12pt,preprint]{aastex}

\newcommand{\lopt}{\ifmmode L_{2500} \else $~L_{2500}$\fi}
\newcommand{\loglopt}{\ifmmode{\rm log}~L_{2500} \else log$~L_{2500}$\fi}
\newcommand{\logz}{\ifmmode{\rm log}~z \else log$~z$\fi}
\newcommand{\ew}{\ifmmode{W_{\lambda}} \else $W_{\lambda}$\fi}
\newcommand{\ax}{\ifmmode{\alpha_x} \else $\alpha_x$\fi} 
\newcommand{\aox}{\ifmmode{\alpha_{\rm ox}} \else $\alpha_{\rm ox}$\fi} 
\newcommand{\fcgs}{\ifmmode erg~cm^{-2}~s^{-1}\else erg~cm$^{-2}$~s$^{-1}$\fi}
\newcommand{\lcgs}{\ifmmode erg~~s^{-1}\else erg~s$^{-1}$\fi}
\newcommand{\kms}{\ifmmode~{\rm km~s}^{-1}\else ~km~s$^{-1}~$\fi}
\newcommand{\mone}{\ifmmode ^{-1}\else$^{-1}$\fi}
\newcommand{\mtwo}{\ifmmode ^{-2}\else$^{-2}$\fi}
\newcommand{\lapprox }{{\lower0.8ex\hbox{$\buildrel <\over\sim$}}}
\newcommand{\gapprox }{{\lower0.8ex\hbox{$\buildrel >\over\sim$}}}
\newcommand{\nh}{\ifmmode{\rm N_{H}} \else N$_{H}$\fi}
\newcommand{\nhgal}{\ifmmode{ N_{H}^{Gal}} \else N$_{H}^{Gal}$\fi}
\newcommand{\nhintr}{\ifmmode{ N_{H}^{intr}} \else N$_{H}^{intr}$\fi}
\newcommand{\nhtot}{\ifmmode{ N_{H}^{tot}} \else N$_{H}^{tot}$\fi}
\newcommand{\atoms}{\ifmmode{\rm ~atoms~cm^{-2}} \else ~atoms cm$^{-2}$\fi}
\newcommand{\cmsq}{\ifmmode{\rm ~cm^{-2}} \else cm$^{-2}$\fi}

\shorttitle{Chandra discovery of a z=4.93 Quasar}
\shortauthors{Silverman et al.}

\begin{document}


\title{Discovery of a z=4.93, X-ray selected quasar by
the Chandra Multiwavelength Project (ChamP)}


\author{John D. Silverman\altaffilmark{1,2,3},
Paul J. Green, Dong-Woo Kim, 
Belinda J. Wilkes, Robert A. Cameron, David Morris, Anil Dosaj\altaffilmark{2}}
\affil{Harvard-Smithsonian Center for Astrophysics, 60 Garden Street,
Cambridge, MA 02138}

\author{Chris Smith}
\affil{Cerro Tololo Inter-American Observatory, National Optical
Astronomy Observatory, Casilla 603, La Serena, Chile}

\author{Leopoldo Infante\altaffilmark{4}}
\affil{Departamento de Astronom\'\i a y Astrof\'\i sica, P.~Universidad Cat\'olica, Casilla 306, Santiago, Chile}

\author{Paul S. Smith}
\affil{Steward Observatory, The University of Arizona, Tucson, AZ 85721}

\author{Buell T. Jannuzi}
\affil{National Optical Astronomy Observatory, P.O. Box 26732, Tucson,
AZ, 85726-6732} 

\author{Smita Mathur}
\affil{Astronomy Department, Ohio State University, 140 West 18th
Avenue, Columbus, OH 43210} 

\altaffiltext{1}{Astronomy Department, University of Virginia,
P.O. Box 3818, Charlottesville, VA, 22903-0818}

\altaffiltext{2}{Visiting Astronomer, Cerro Tololo Inter-American Observatory,
National Optical Astronomy Observatory, which is operated by the
Association of Universities for Research in Astronomy, Inc. (AURA)
under cooperative agreement with the National Science Foundation.}

\altaffiltext{3}{email: jds@head-cfa.harvard.edu}
\altaffiltext{4}{Visiting astronomer, ESO New Technology Telescope}


\begin{abstract}

We present X-ray and optical observations of CXOMP\,J213945.0-234655, a
high redshift ($z=4.93$) quasar discovered through the Chandra
Multiwavelength Project (ChaMP).  This object is the most distant
X-ray selected quasar published, with a rest-frame X-ray luminosity of
L$_{X}=5.9\times 10^{44}$ erg~s$^{-1}$ (measured in the 0.3--2.5~keV
band and corrected for Galactic absorption).  CXOMP\,J213945.0-234655
is a $g^{\prime}$ dropout object ($>26.2$), with $r^{\prime}=22.87$
and $i^{\prime}=21.36$.  The rest-frame X-ray to optical flux ratio is
similar to quasars at lower redshifts and slightly X-ray bright
relative to $z>4$ optically-selected quasars observed with Chandra.
The ChaMP is beginning to acquire significant numbers of high redshift
quasars to investigate the X-ray luminosity function out to $z\sim5$.

\end{abstract}




\section{Introduction}

The observed characteristics of known quasars are remarkably similar
over a broad range of redshift.  For example, X-ray studies utilizing
the ROSAT database \citep{gr95,ka00}, show little variation of the ratio of
X-ray to optical flux for optically selected quasars.  Also, the rest
frame UV spectra of quasars, including the broad Ly$\alpha$, NV and
CIV emission lines, are nearly identical for a large range of redshift
and present no evidence for subsolar metallicities even up to a
$z\sim6$ \citep{fa01}.

Even though the individual properties of quasars are similar, the
co-moving space density of quasars changes drastically with redshift.
At high redshift ($z>4$), a significant dropoff in the co-moving space
density of quasars seen in optical (e.g., Schmidt et al. 1995; Warren
et al. 1994; Osmer 1982) and radio surveys \citep{sh96} hints at
either the detection of the onset of accretion onto supermassive black
holes or a missed high-redshift population, possibly due to
obscuration.  X-ray selected quasars from ROSAT have been used to
support the latter interpretation based on evidence for constant space
densities beyond a redshift of 2 \citep{mi00}.  Unfortunately, the
ROSAT sample size is small with only 8 quasars beyond a redshift of 3.

Significant numbers of quasars with $z>4$ are being amassed to
investigate both their intrinsic properties and the evolutionary
behavior of the quasar population. The Sloan Digital Sky Survey (SDSS)
reports 123 optically selected quasars with $z>4$
\citep{sc01,an01}.  However, optical surveys suffer from selection
effects due to intrinsic obscuration and the intervening Ly$\alpha$
forest.  Current X-ray surveys with Chandra and XMM do not have a
strong selection effect based on redshift and can detect emission up
to 10~keV (observed frame) to reveal hidden populations of active
galactic nuclei (AGN) including heavily obscured quasars
\citep{no01,st01}.  High-$z$ objects can be detected through
a larger intrinsic absorbing column of gas and dust because the
observed-frame X-ray bandpass corresponds to higher energy, more
penetrating X-rays at the source.\footnote{The observed-frame,
effective absorbing column is $N_{\rm H}^{\rm eff}\sim N_{\rm
H}/(1+z)^{2.6}$ (Wilman \& Fabian 1999).}  Therefore, optical and
X-ray surveys will complement each other, providing a fair census of
mass accretion onto black holes at high redshift.

Larger samples of X-ray observations of $z>4$ quasars are needed since
there are currently only 24 \citep{vi01}, of which only 3 are X-ray
selected quasars. Chandra and XMM-Newton are beginning to probe faint
flux levels for the first time to detect the high-$z$ quasar
population.  Initial Chandra and XMM-Newton observations of optically
selected quasars have shown a systematically lower X-ray flux relative
to the optical at high redshift \citep{vi01,br01a}.

In this paper, we present the X-ray and optical properties of a newly
discovered, X-ray selected $z=4.93$ quasar with the Chandra
Observatory.  This quasar is the highest redshift object
published\footnote{A $z\sim5.2$, X-ray selected
quasar detected in the CDF-N was presented at the 199th AAS meeting
(Brandt 2001b).} from an X-ray survey.

These results are a component of the Chandra Multiwavelength Project
(ChaMP; Wilkes et al. 2001).  A primary aim of the ChaMP is to measure
the intrinsic luminosity function of quasars and lower luminosity AGN
out to $z\sim5$. The survey will provide a medium-depth, wide-area
sample of serendipitous X-ray sources from archival Chandra fields in
Cycles~1 and 2 covering $\sim 14$ deg$^2$.  The broadband sensitivity
between 0.3--8.0 keV enables the selection to be far less affected by
absorption than previous optical, UV, or soft X-ray surveys.
Chandra's small point spread function ($\sim$1$\arcsec$ resolution
on-axis) and low background allow sources to be detected to fainter
flux levels, while the $\sim 1^{\prime\prime}$ X-ray astrometry
greatly facilitates unambiguous optical identification of X-ray
counterparts.  The project will effectively bridge the gap between
flux limits achieved with the Chandra deep field observations and
those of past ROSAT surveys.

Throughout this paper, we assume H$_{\circ}$=50 km s$^{-1}$
Mpc$^{-1}$ and a flat cosmology with q$_{\circ}$=0.5.

\section{Observations and data analysis}

\subsection{X-ray}

The X-ray source CXOMP\,J213945.0-234655 (Seq. 800104) was observed on
November 18, 1999 by Chandra \citep{we00} with the Advanced CCD
Imaging Spectrometer (ACIS-I; Nousek et al. 1998) in the field of the
X-ray cluster MS~2137.3-2353 (M. Wise, PI).  We have used data
reprocessed (in April 2001) at CXC\footnote{CXCDS versions
R4CU5UPD14.1, along with ACIS calibration data from the Chandra
CALDB~2.0b.}. We then ran a detection algorthim XPIPE
\citep{ki02} which was specifically designed for the ChaMP to produce
a uniform and high quality source catalog.

\begin{figure*}
\epsscale{0.8}
\plotone{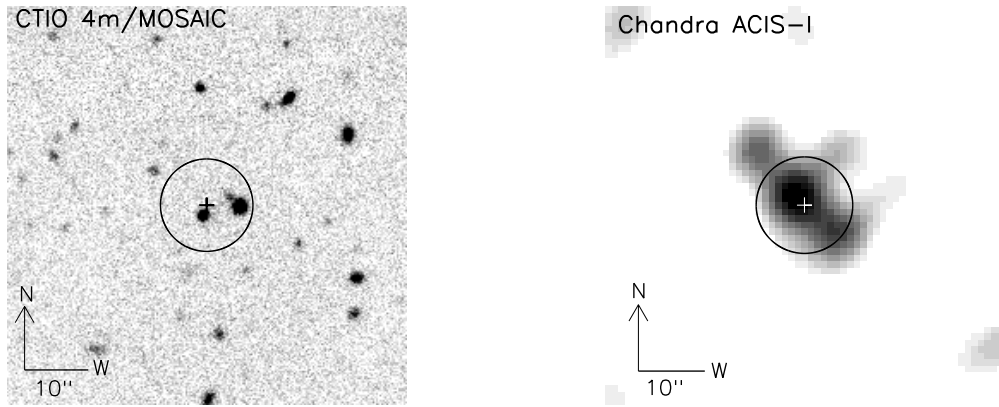}
\caption{Optical ($i^{\prime}$) and X-ray (0.3--2.5 keV) imaging.  
To improve the visual clarity, in this figure we have smoothed the
Chandra image with a gaussian function ($\sigma$=1.5$\arcsec$). The
spatial distribution of the 17 X-ray counts at $9.1\arcmin$ off-axis is as
expected from a point source. The black circle shows the region
containing 50\% of the encircled energy (radius=7.3$\arcsec$) of the
Chandra counts. The cross marks the centroid of the X-ray emission in
both images.\label{image}}
\end{figure*}

CXOMP\,J213945.0-234655 is one of 72 sources detected using
CIAO/{\tt{wavdetect}} (Freeman et al.\ 2002) within the ACIS
configuration (Figure~\ref{image}).  The 41~ksec observation yielded a
net 16.7$\pm$7.5 counts within the soft bandpass (0.3--2.5 keV) and no
counts in the hard bandpass (2.5--8.0 keV).  This corresponds to a
Galactic absorption corrected, observed frame X-ray flux of
$f(0.3-2.5$ keV)=2.82 $\pm1.26\times 10^{-15}$ \fcgs (Table 1).

The source naming convention of the ChaMP
(CXOMP\,Jhhmmss.s$\pm$ddmmss) is given with a prefix CXOMP (Chandra
X-ray Observatory Multiwavelength Project) and affixed with the
truncated J2000 position of the X-ray source after a mean field offset
correction is applied, derived from the cross-correlation of optical
and X-ray sources in each field.

\subsection{Optical Imaging and source matching}

We obtained optical imaging of the field in three NOAO/CTIO SDSS
filters ($g^{\prime}$,$r^{\prime}$ and $i^{\prime}$; Fukugita et
al. 1996) with the CTIO 4m/MOSAIC on October 29, 2000 as part of the
ChaMP optical identification program \citep{gr02}.  Integration time
in each band ranged from 12--15 minutes during seeing of
1.3$\arcsec$--1.8$\arcsec$ FWHM. Image reduction was performed with the
IRAF(v2.11)/MSCRED package.  We used SExtractor \citep{be96} to detect
sources, and measure (pixel) positions and magnitudes.  Landolt standard
stars were transformed to the SDSS photometric system \citep{fu96} and
used to calibrate the photometric solution. Following the convention
of the early data release of the SDSS quasar catalog
\citep{sc01}, we present the optical photometry here as $g^{\ast}$,
$r^{\ast}$ and $i^{\ast}$ since the SDSS photometry system is not yet
finalized and the CTIO filters are not a perfect match to the SDSS
filters.  The limiting magnitudes for a point source are given as the
mean of 3$\sigma$ detections: $g^{\ast}$=26.18, $r^{\ast}$=25.54,
$i^{\ast}$=25.11.

As evident from Figure~\ref{image}, there are three optical sources
detected down to a limiting $i^{\ast}$ magnitude of 25.1 within the
50\% encircled energy radius of the X-ray centroid.  The two primary
candidates, based on optical brightness, have offsets between the
optical and X-ray positions of 1.87$\arcsec$ and 4.94$\arcsec$.  To
determine whether either of these sources are the likely counterpart
to the X-ray detection, we have determined errors associated with the
X-ray astrometric solution.

\citet{ki02} have carried out extensive simulations of point sources
generated using the SAOSAC raytrace program
(http://hea-www.harvard.edu/MST/) and detected using
CIAO/{\tt{wavdetect}}.  For weak sources of $\sim 20$ counts between
8$\arcmin$--10$\arcmin$, off-axis from the aim point, the reported
X-ray centroid position is correct within $1.8\arcsec$, corresponding
to a $1\sigma$ confidence contour.  Therefore, the nearby optical
source ($\Delta$r=1.9$\arcsec$) is the likely counterpart to the X-ray
detection.  The (J2000) position of the optical counterpart as
measured from the $r^{\prime}$ image referenced to the Guide Star
Catalog II\footnote{The Guide Star Catalogue-II is a joint project of
the Space Telescope Science Institute and the Osservatorio Astronomico
di Torino.} is
$\alpha=21^{h}~39^{m}~44.99^{s}~\delta=-23^{\circ}~46{\arcmin}~56.6{\arcsec}$.

\subsection{Optical spectroscopy}

\begin{figure*}
\epsscale{0.8}
\plotone{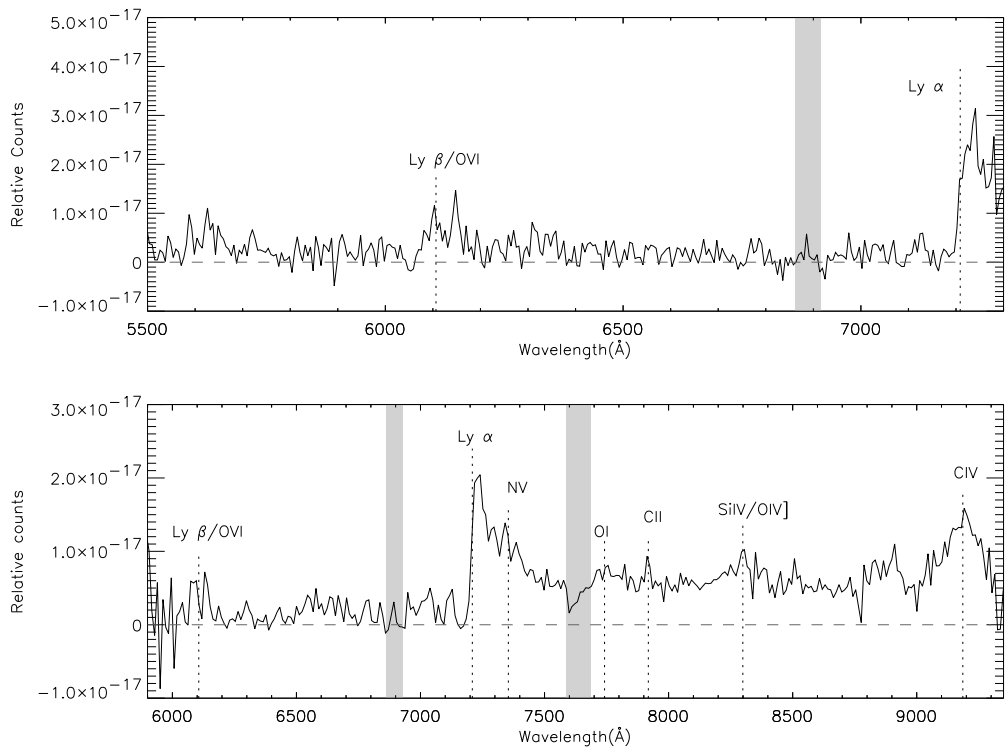}
\caption{Optical Spectroscopy of CXOMP\,J213945.0-234655.  The top
spectrum is the discovery observation taken with the CTIO 4m/HYDRA on
October 15, 2001.  The spectrum has been binned to produce a
resolution of 16.4 \rm{\AA}.  The bottom figure is a followup
observation with the NTT on the next evening to detect spectral
features redward of Ly$\alpha$ (11 \AA~resolution). Dashed lines
indicate the expected positions of emission lines at a redshift of
4.93.  Shaded regions mark the uncorrected telluric O$_{2}$ absorption
band regions.
\label{spectra}}  
\end{figure*}

We obtained a low resolution optical spectrum of
CXOMP\,J213945.0-234655 (Figure~\ref{spectra}) with the CTIO 4m/HYDRA
multi-fiber spectrograph on October 15, 2001.  Spectra of 17 of 22
optical counterparts to X-ray sources with a magnitude $r^{\ast}<21$
were acquired in a three hour integration within the Chandra field.
The spectrograph has 2$\arcsec$ diameter fibers and was configured
with a 527 $l$/mm grating that provided $\sim$2800 \rm{\AA} of
spectral coverage with a resolution of $\sim$4 \rm{\AA}.  The sky
background was measured using 81 fibers not assigned to the Chandra
X-ray detections within the 1$^{\circ}$ field spectrograph.  We
processed the data using the IRAF(v2.11)/HYDRA reduction package.

An additional spectrum of the high redshift quasar
(Figure~\ref{spectra}) and the optically brighter source 4.9\arcsec\,
west of the Chandra X-ray position were obtained on the following
evening with the ESO/NTT 3.5m to verify the intriguing Hydra spectrum
and obtain greater wavelength coverage.  A 300 $l$/mm grating was
implemented with a wavelength coverage of 4000
\rm{\AA} and a resolution of $\sim$11 \rm{\AA}. Due to poor weather
conditions at the end of the evening, flux calibration was done using
the standard star LTT 2415 observed the following night.  From the
NTT spectrum, we classify the brighter object as an M3 dwarf with no
sign of emission lines, confirming the quasar as the optical
counterpart of the X-ray source.

We measured a mean redshift $z$=4.930 $\pm$0.004 from the
Ly$\beta$+OVI, CII, SiIV+OIV] and CIV emission lines in the NTT
spectrum.  Using this redshift, the Ly$\alpha$ line centroid is
shifted by $\approx$4 \rm{\AA} redward from the expected rest
wavelength, probably due to significant HI absorption.  This is
similar to the mean shift of Ly$\alpha$ observed in a sample of 33
high redshift quasars by \citet{sc91}.

The spectrum obtained at the NTT was used to measure the rest-frame
equivalent widths of Ly$\beta$/OVI ($30\pm7$ \AA), Ly$\alpha$+NV
($73\pm5$ \AA), and CIV ($40\pm8$ \AA). For comparison, we also
measured Ly$\alpha$+NV for 10 high redshift quasars in the range
$4.8<z<5.1$ from the SDSS spectra of Anderson et al. (2001).  This
subsample has a similar mean redshift (4.91), but with an average
$i^*=19.7$ is 4.5x more optically luminous than
CXOMP\,J213945.0-234655.  Nevertheless, the mean rest-frame equivalent
width of Ly$\alpha$+NV in the SDSS subsample is consistent at 79 \AA,
with an RMS dispersion of 27 \AA.  The poor S/N of the SDSS spectra
and the strong Ly$\alpha$ forest, prevent meaningful comparison of
other line strengths.

\section{Results}             

To compare the broad band spectral energy distribution of
CXOMP\,J213945.0-234655 to other X-ray detected quasars, we have
calculated $\alpha$$_{ox}$ \citep{ta79}, the slope of a hypothetical
powerlaw between the X-ray and optical flux.  The rest-frame,
monochromatic luminosity at 2 keV corresponding to the derived X-ray
flux is log $l_{2~keV}=26.76$ erg~s$^{-1}$~Hz$^{-1}$.  Assuming
$\alpha$=0.5 for the optical continuum powerlaw slope, we derive the
rest-frame, monochromatic optical luminosity at 2500
\AA\, from the $i^{\ast}$ magnitude to be log $l_{2500\AA}=30.73$ erg
s$^{-1}$Hz$^{-1}$.  We thus find $\alpha_{ox}=1.52^{+0.08}_{-0.05}$.
Table 1 lists the measured X-ray and optical properties of
CXOMP\,J213945.0-234655.

\begin{figure*}
\epsscale{1.0}
\plotone{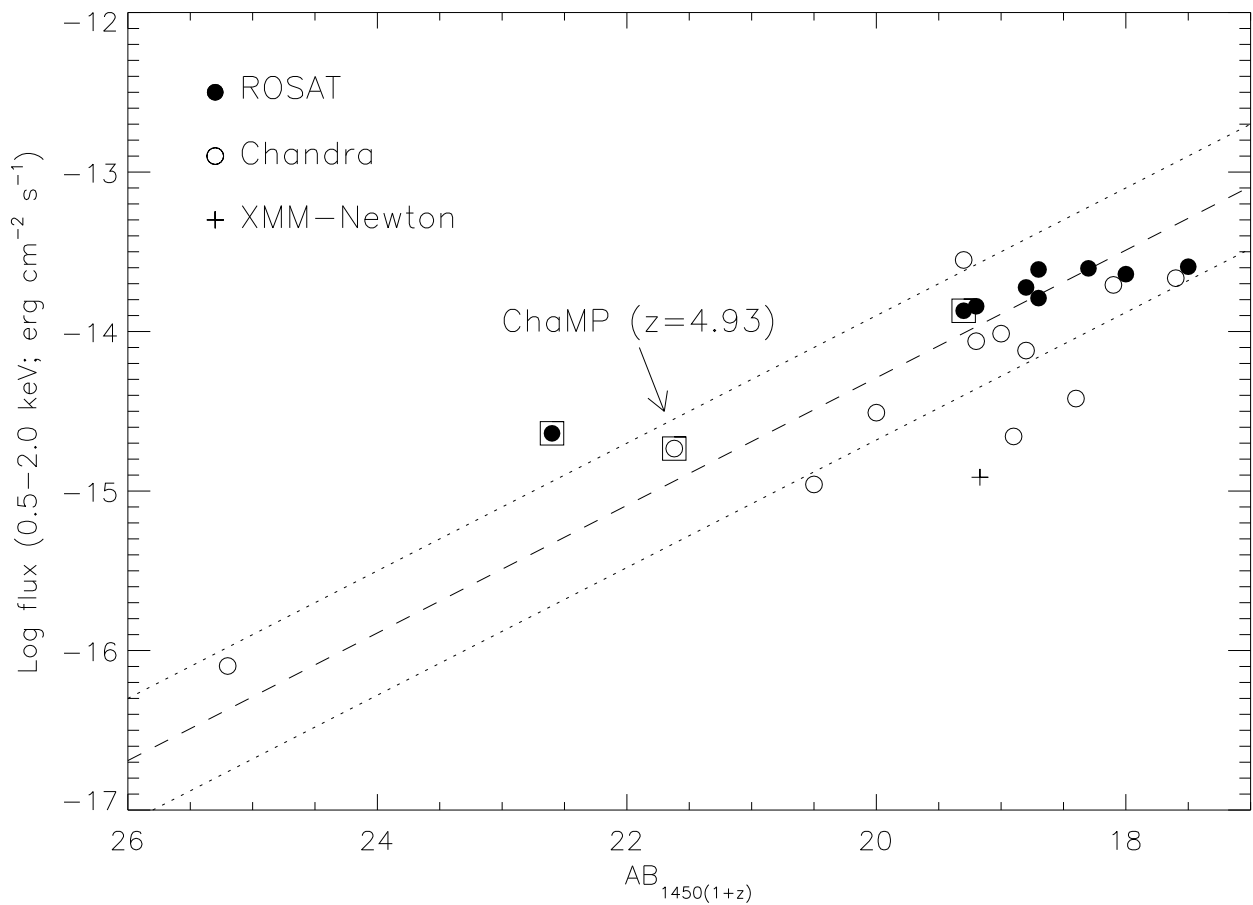}
\caption{X-ray to optical flux correlation for $z>4$ AGN (adapted
from Vignali et al. 2001).  The primary symbols represent the X-ray
observatory used.  Squares mark X-ray selected AGN.  The faintest
source shown is a radio-selected Seyfert galaxy at $z=4.424$
\citep{br01c}. The dashed lines displays the relation for AGN with
$\alpha_{ox}$=1.6 $\pm$0.15 at $z$=4.9 \citep{gr95}.\label{aox}}
\end{figure*}

We compare the X-ray to optical flux ratio of CXOMP\,J213945.0-234655
to other $z>4$ quasars by plotting the observed-frame, Galactic
absorption corrected 0.5--2.0 keV X-ray flux versus the
AB$_{1450(1+z)}$ magnitude (Figure~\ref{aox}).  The plotted lines
represent the locus of points for a hypothetical quasar with a wide
range of luminosities and an $\alpha_{ox}$=1.6 $\pm$0.15 \citep{gr95},
representative of the mean for quasars selected from the Large Bright
Quasar Survey and detected in the ROSAT All-Sky Survey.  The
$\alpha_{ox}$ of CXOMP\,J213945.0-234655 is comparable with low
redshift quasars in contrast to the X-ray faint Chandra detections of
optically selected quasars at $z>$4 \citep{vi01}.  The X-ray weakness
of the latter may be due to intrinsic absorption by large amounts of
gas in the quasars' host galaxies.

X-ray and optical observations of CXOMP\,J213945.0-234655 show no
direct evidence of significant obscuration.  The optical color
($r^{\ast}-i^{\ast}=1.51\pm0.12$) is consistent with optically
selected quasars.  We measured the mean color $r^{\ast}-i^{\ast}$ from
15 SDSS quasars \citep{an01} with $4.7<z<5.2$ to be 1.69 with RMS
dispersion of 0.30.  The upper limit to the X-ray hardness ratio
($<-0.54$) hints at an unobscured X-ray spectrum, although a
moderately absorbed component, if present, would be redshifted out of
the Chandra bandpass.

X-ray selected samples may be less biased against absorbers (both
intrinsic and line-of-sight) than are optically-selected samples, an
advantage expected to be especially important at high redshifts.  From
our flux-calibrated NTT spectrum, we measure
$D_A=(f_{cont}-f_{obs})/f_{cont}$, the flux decrement caused by the
Ly$\alpha$ forest (Oke \& Korycansky 1982) relative to an extrapolated
power-law continuum\footnote{As in Fan et al. (2001), we measure the
observed flux $f_{obs}$ relative to a $f_{\lambda}\propto
\lambda^{-1.5}$ power-law continuum normalized to the observed flux in
the region $1270\pm10$\AA\, in the rest-frame. We derive uncertainties
by measuring against continua with slopes in the range
$-0.5<\alpha\le1.5$.} in the region between rest-frame limits $1050 -
1170$ \AA.  The value we measure of $D_A=0.79\pm0.02$ is between the
$\overline{z}\sim4$ measurement of 0.54 from Rauch et al. (1997) and
the $z\sim6$ measurements of $D_A\sim0.9$ from 4 SDSS quasars in
Becker et al. (2001).  While CXOMP\,J213945.0-234655 thus appears
consistent with the handful of bracketing measurements of
optically-selected quasars (see also Riediger et al. 1998), more high
redshift X-ray selected quasars are needed to test possible biases
caused by absorption.

CXOMP\,J213945.0-234655 exemplifies the potential for the ChaMP
project to detect quasars with fluxes at the faint end of the
$f_{x}-f_{opt}$ parameter space (Figure~3).  This will allow the ChaMP
to acquire significant numbers of high redshift quasars.  From the
first year of spectroscopic followup of Chandra X-ray sources to
$i^{\prime}\lapprox 21$, we currently have 22 newly identified quasars
with $z>2$ and eight with $z>3$, approximately 2--3 such objects per field.
Nearly 5\% of ChaMP sources identified to date are $z>3$ quasars.

\section{Conclusion}

We present the discovery of CXOMP\,J213945.0-234655, at $z=4.93$ the
most distant X-ray selected object published to date.  With a measured
optical to X-ray flux ratio $\alpha_{ox}$=1.52,
CXOMP\,J213945.0-234655 is similar to low redshift quasars, in
contrast to several optically-selected $z>4$ quasars previously
detected by Chandra.

This detection highlights the importance of wide area, intermediate
depth surveys like the ChaMP for studies of the high redshift quasar
population ($z\sim$ 3 to 5).  The
ChaMP\footnote{http://hea-www.harvard.edu/CHAMP/} has begun to amass a
sample of high redshift, X-ray selected quasars with the goal of
measuring the cosmic evolution of accretion-powered sources relatively
unhampered by the absorption and reddening that affects optical
surveys.

\acknowledgments
We gratefully acknowledge support for this Chandra archival research
from NASA grant AR1-2003X.  RAC, AD, PJG, DK, DM, and BW also
acknowledge support through NASA Contract NASA contract NAS8-39073
(CXC).  LI is grateful to ``Proyecto Puente PUC'' and Center for
Astrophysics FONDAP for partial financial support.  BTJ acknowledges
research support from the National Science Foundation, through their
cooperative agreement with AURA, Inc., for the operation of the NOAO.
We are thankful to Sam Barden and Tom Ingerson (NOAO) for building and
commissioning Hydra/CTIO.  We greatly appreciate the observing support
from Knut Olsen (NOAO), and constructive comments by Harvey Tananbaum and
Dan Harris.

\begin{deluxetable}{llll}
\tabletypesize{\scriptsize}
\tablewidth{10cm}
\tablecaption{Properties of CXOMP\,J213945.0-234655}
\tablehead{
\colhead{Parameter} &\colhead{Value}& \colhead{Parameter} &\colhead{Value}
}
\startdata
$\alpha\rm_{J2000}$\tablenotemark{1}&21~39~44.99&X-ray counts\tablenotemark{4}&16.7 $\pm$ 7.5\\
$\delta\rm_{J2000}$\tablenotemark{1}&$-23~46~56.6$&f$_{X}$(erg s$^{-1}$ cm$^{-2}$)\tablenotemark{3,4}  &(2.82 $\pm$1.26) x 10$^{-15}$\\
z &4.930 $\pm$ 0.004&L$_{X}$(erg s$^{-1}$)\tablenotemark{3,5}&(5.89 $\pm$2.63) x 10$^{44}$\\
g* &$>$26.2&Hardness Ratio\tablenotemark{6} &$<-0.54$\\
r* &22.87 $\pm$0.07&$\alpha_{ox}$ &1.52$^{+0.08}_{-0.05}$\\
i* &21.36 $\pm$0.10&AB$_{1450(1+z)}$\tablenotemark{2} &21.62\\
\enddata

\tablenotetext{1}{units RA (hms), DEC ($\circ~\prime~\prime\prime$); error $<$0.5$\arcsec$}
\tablenotetext{2}{observed monochromatic, Galactic absorption
corrected, AB$_{1450(1+z)}$ magnitude \citep{fu96} emitted at
1450 \rm{\AA} in the quasar's rest-frame; based on an assumed optical powerlaw spectrum
($S_{\nu}\propto \nu^{-\alpha}; ~\alpha$=0.5)}
\tablenotetext{3}{based on an assumed X-ray powerlaw spectrum
($S_{E}\propto E^{-\alpha}; ~\alpha$=1.0); Galactic absorption
corrected ($N_H=3.55\times 10^{20}$ cm$^{-2}$; Dickey \& Lockman 1990)}
\tablenotetext{4}{observed-frame; 0.3--2.5 keV}
\tablenotetext{5}{rest-frame; 0.3--2.5 keV}
\tablenotetext{6}{(H--S)/(H+S); soft band(S): 0.3-2.5 keV, hard band(H): 2.5-8.0 keV}
\end{deluxetable}

\clearpage

\end{document}